\documentclass[final,3p,times,review]{elsarticle}

\usepackage{amssymb}
\usepackage{multirow}

\journal{Optical Materials}

\begin{document}

\begin{frontmatter}



\title{Orange emission in Pr$^{3+}$-doped fluoroindate glasses}


\author[label1]{Danilo Manzani}
\author[label2,label3]{David Pab{\oe}uf}
\author[label1]{Sidney J. L. Ribeiro}
\author[label3]{Philippe Goldner}
\author[label2]{Fabien Bretenaker}

\address[label1]{Institute of Chemistry-S\~ao Paulo State Univ-UNESP, CP 355, Araraquara-SP, 14801-970, Brazil}
\address[label2]{Laboratoire Aim\'e-Cotton, CNRS-Universit\'e Paris Sud 11, Orsay, France}
\address[label3]{Laboratoire de Chimie de la Mati\`ere Condens\'ee de Paris, UMR 7574 CNRS-Chimie ParisTech, Paris, France}

\begin{abstract}
We synthesize and study the properties of praseodymium doped fluoroindate glasses. Glass compositions with praseodymium molar concentrations up to 5\,\% were obtained with good optical quality. Thermal, optical, and luminescence properties are investigated. Judd-Ofelt analysis is used to determine radiative lifetime and emission cross-section of the orange transition originating from the $^3$P$_0$ level. We find that these glasses are good candidates for the realization of blue diode laser pumped orange lasers for quantum information processing applications.
\end{abstract}


\end{frontmatter}


\section{Introduction}
\label{Introduction}
Rare-earth (RE) doped glasses capable of efficient frequency conversion have received great attention due to the possibilities of using these materials to build solid-state lasers operating in the green-red region, as well as for developing infrared devices such as optical fiber amplifiers usable for telecommunications. Several investigations have been conducted towards the use of RE-doped fluoride glasses to serve these purposes, in particular view of their optical and structural characteristics. They can indeed sustain high concentrations of RE ions, have a low multiphonon absorption and a wide transparency window in the visible to infrared region up to $7\,\mu\mathrm{m}$. ZBLAN is the most important glass of this class of materials \cite{Poulain1975}. However, difficulties such as poor mechanical and chemical stability are commonly encountered when one uses ZBLAN optical fibers. In 1993, Messaddeq, Poulain, and coworkers reported a new InF$_3$-based glass that exhibits improved chemical and mechanical stability with respect to ZBLAN glass \cite{Messaddeq1993}, and can also be fabricated as optical fibers and planar waveguides \cite{DeMelo1995,Bernier2009}. Furthermore, owing to the small multiphonon emission rate in this glass, the non-radiative relaxation rate between close levels is reduced and some RE fluorescence transitions, which are not observed in ZBLAN glass, can be active in this host \cite{Mazurak1984}. Among RE ions, praseodymium ion (Pr$^{3+}$) has many possible applications owing to its large number of absorption bands in the UV, visible and near infrared, that provide the possibility for simultaneous emission in the blue, green, orange, red, and infrared regions \cite{Kaminskii1990}. Pr$^{3+}$-doped glasses and crystals are presently developed for applications to optical amplifiers, up-converters, and opto-electronic devices \cite{Moorty2005,Sharma2000,Ratnakaram2004,Florez1997}, in particular for fiber-based optical communication systems operating with 1.3 $\mu$m radiation \cite{Wei1995}. More generally, fluoroindate glasses are promising materials for various photonics applications in the visible and near-infrared domains. There  is  an  increasing  interest  in  the  determination  of  the  optical  properties  of  heavy  metal fluoride  glasses  doped  with  rare-earth  ions. Some devices with excellent optical characteristics have been reported by using ZrF$_4$ \cite{Nazabal2012} and  InF$_3$-based glasses \cite{Adam1988,Kazuyuki1992}. Therefore, our aim here is to report the results of an investigation of the optical and fluorescence properties of Pr$^{3+}$-doped InF$_3$-ZnF$_2$-BaF$_2$-SrF$_2$ glasses. In particular, we put special emphasis on the question to know whether such glasses are good candidates for the realization of orange lasers using the same transition as recently demonstrated lasers based on Pr$^{3+}$ doped fluoride crystals \cite{My2008,Paboeuf2011}. Among other applications, such lasers could be used for the coherent driving of  rare-earth ions used for quantum information processing \cite{Longdell2004,Hedges2011}.

\section{Experimental}

\begin{table*}[]
\caption{\label{table1} Prepared Pr$^{3+}$-doped ISZB glass samples.}
\centering
\begin{tabular}{c|c|c} 
   \hline
  Pr$^{3+}$ concentration & Glass composition & Sample label \\
  (mol \%) & (mol \%) & \\
    \hline
    \hline
    0.05 & \multirow{6}{*}{40 InF$_3$-- 20 ZnF$_2$-- 20 SrF$_2$-- 20 BaF$_2$} & IZSB005\\
    0.1 & & IZSB01\\
    0.2 & & IZSB02\\
    0.5 & & IZSB05\\
    1.0 & & IZSB1\\
     5.0 & & IZSB5\\
    \hline
\end{tabular}
\end{table*}

\subsection{Glass synthesis}
The glass samples were synthesized using the conventional melt-casting method. The starting powdered materials were indium fluoride InF$_3$, zinc fluoride ZnF$_2$ (3N), strontium fluoride SrF$_2$ (3N), barium fluoride BaF$_2$ (3N), and praseodymium fluoride PrF$_3$. In a first step, the powders were weighed in order to obtain a 6~g glass bulk whose molar composition is $(100-x)$[40InF$_3$-20ZnF$_2$-20SrF$_2$-20BaF$_2$] : $x$PrF$_3$ ($x = 0.05, 0.1, 0.2, 0.5, 1.0, \mathrm{and }\ 5.0$). The powders were then thoroughly mixed and loaded in a platinum crucible. An excess of ammonium bifluoride, NH$_4$HF$_2$, was added in all compositions to reduce the amount of species produced from oxidation reaction with environmental water adsorbed in precursor powders. The mixture was heated at $350\,^{\circ}$C during 1 h for fluorination reaction and then melted at $920\,^{\circ}$C for 30 minutes in an electrical furnace. This ensured the complete elimination of NH$_3$ and HF from the decomposition of ammonium bifluoride and a good homogenization and fining. Finally, the melt was cooled down in a stainless mold pre-heated at $20\,^{\circ}$C below the glass transition temperature, annealed at this temperature for 2 h and slowly cooled down to room temperature to minimize residual internal stress. In the following, for the sake of readability, the host glass is labeled IZSB and the samples studied in this work are presented in Table \ref{table1}. The Pr$^{3+}$-doped IZSB glass samples of $30\times20\times2\ \mathrm{mm}^3$ dimensions (length, width and thickness, respectively) of very good optical quality were finally polished for optical measurements.

\subsection{Thermal properties}
The glass transition temperature, $T_g$, and the thermal stability parameter against devitrification ($\Delta T = T_x - T_g$), where $T_x$ is the crystallization temperature, are commonly used to estimate the thermal stability of the glasses \cite{Dietzel1968}. We measured them using differential scanning calorimetry in the 200 to 600$^{\circ}$C temperature range under N$_2$ atmosphere at a heating rate of 10$^{\circ}$C/min, using a TA Instruments DSC 2910 calorimeter, with a maximum error of $\pm2^{\circ}$C for $T_g$ and $T_x$. The corresponding results are summarized in Table \ref{table2}.

\begin{table*}[]
\caption{\label{table2} Glass samples characteristic temperatures and refractive indices in the visible range.}
\centering
\begin{tabular}{c|c|c|c|c|c} 
\hline Glasses & $T_g\  (^{\circ}\mathrm{C})$  & $T_x\  (^{\circ}\mathrm{C})$ & $\Delta T\  (^{\circ}\mathrm{C})$ & \multicolumn{2}{c}{Refractive index}\\
\hline
 & & & & 532.4 nm & 632.8 nm \\
     \hline
    \hline
    IZSB0 & \multirow{4}{*}{287}& \multirow{4}{*}{368}& \multirow{4}{*}{81} & 1.4951 & 1.4923\\
    IZSB01 & & & & 1.4958 & 1.4930\\
    IZSB02 & & & & 1.4967 & 1.4937\\
    IZSB05 & & & & 1.4976 & 1.4946\\
     \hline
     ZBLAN \cite{Adam2001} & 262 & 352 & 90 & 1.5 & 1.5\\
     \hline
\end{tabular}
\end{table*}

\subsection{Optical and luminescence measurements}
The absorption spectra measurements were performed on the polished IZSB:Pr glasses of about 2 mm thickness by using a Cary 500 spectrophotometer (Varian) at room-temperature from 200 to 3000~nm. Emission spectra were recorded by using a Horiba Jobin Yvon fluorimeter equipped with a photomultiplier tube (PMT) sensitive from 250 to 800 nm and a Xe lamp (450 W) operating at 480 nm with continuous excitation was utilized as the excitation source. The emission was measured at $30\,^{\circ}$ from the excitation beam. Slits were adjusted to lead to a resolution of 2 and 4 nm for excitation and emission, respectively. All measurements have been performed at room temperature and corrected by the instrument response.

The refractive indices of some of the glasses were determined by measuring the critical angle for the sample/prism interface (accuracy ±0.0001), using three laser beam wavelengths, 542, 633, and 1550 nm (Metricon-2010 instrument).

Lifetime measurements of $^3$P$_0$ praseodymium excited level were carried out using a pulsed EKSPLA optical parametric oscillator (model \# NT342B) to excite the glasses. Excitation was carried out at 480~nm along the $^3$H$_4\to^3$P$_0$. Fluorescence  from the $^3$P$_0$ level to the $^3$H$_6$ ground state was detected at 604~nm with a Jobin-Yvon HR250 monochromator and a Princeton Instrument intensified CCD or a photomultiplier tube. 

\begin{figure}[h]
      \includegraphics[width=0.5\linewidth]{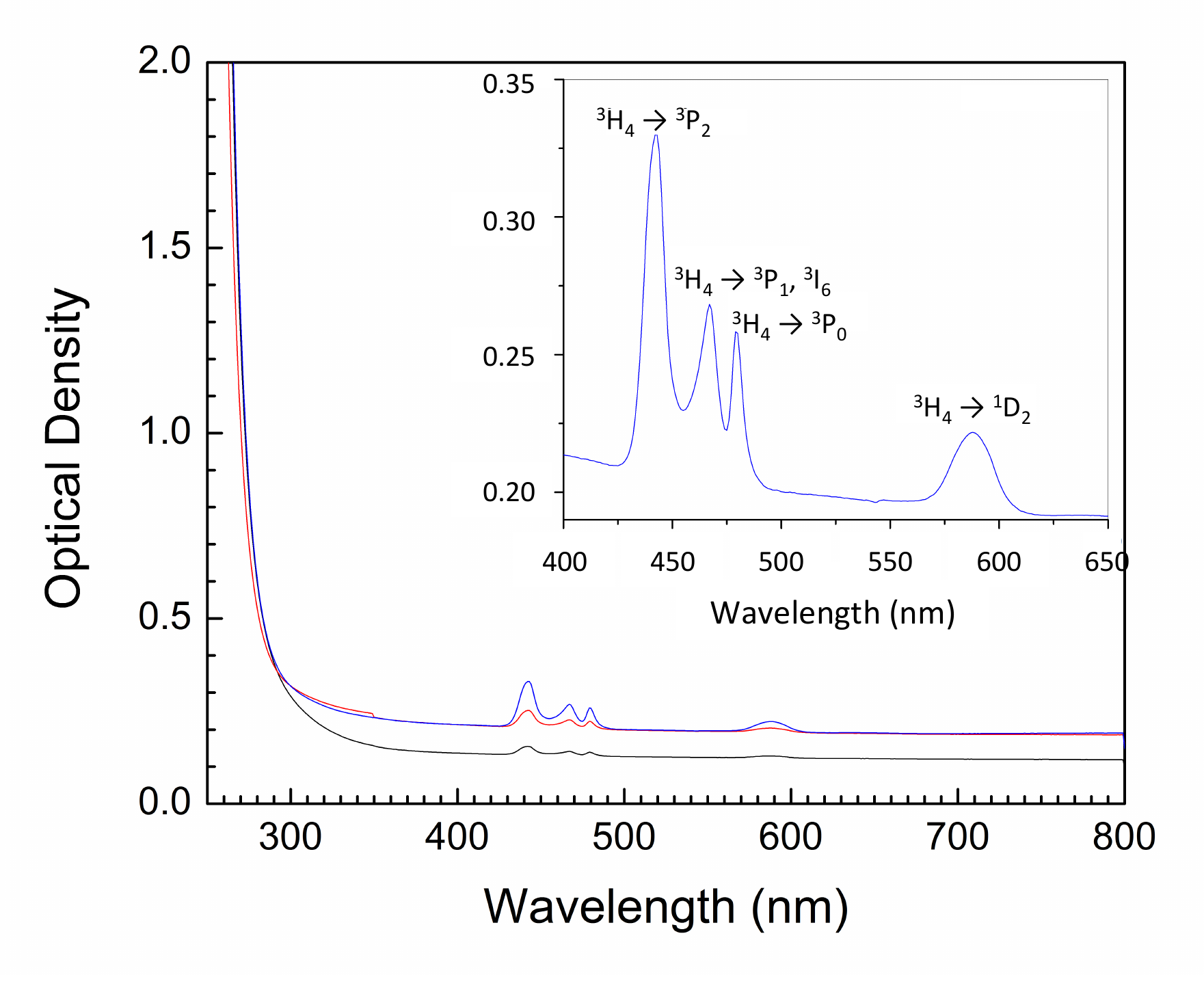}
  \caption{UV and visible absorption spectra of some of the Pr$^{3+}$-doped IZSB glass samples. Black line: IZSB01, red line: IZSB02, blue line: IZSB05. Inset: zoom on the absorption transitions from the ground state of Pr$^{3+}$ ion for IZSB05 (sample thickness is 2 mm).}\label{fig1}
\end{figure}

\section{Results and discussion}
Homogenous and slightly greenish glasses samples up to 2~mm thick were obtained in the pseudo-quaternary system InF$_3$-ZnF$_2$-(SrF$_2$-BaF$_2$) doped with 0.05, 0.1, 0.2, 0.5, 1.0, and 5.0\% mol of Pr$^{3+}$ ions. No bubbles or crystals were observed in the glass bulk, which is very important when the objective is to use these glasses in optical devices that cannot bear large scattering losses. The typical amorphous halo was observed by X-ray diffraction for all compositions. Table \ref{table2} displays the characteristic temperatures, the thermal stability parameter ($\Delta T=T_x-T_g$) and the refractive indices in the visible range (532.4 and 632.8 nm) measured for some of the compositions. 

When the doping concentration of Pr$^{3+}$ ions varies, the glass transition temperature and thermal stability parameter remain unchanged. The value of the thermal stability parameter against devitrification is comparable with the ones of the most common ZBLAN glasses \cite{Adam2001}, making the InF$_3$-based glasses good potential candidates for optical fiber production.

Contrary to the behavior of the characteristic temperatures, a slight increase of the value of the linear refractive index $n_0$ with the praseodymium ion concentration was observed. This index ranged from 1.4951 to 1.4976 at 532.4~nm, and 1.4923 to 1.4946 at 632.8~nm, for Pr$^{3+}$ ion concentration ranging from 0 to 0.5\% (see Table \ref{table2}). The increase of $n_0$ can be easily understood by the high polarizability of Pr$^{3+}$ ions owing to their extended electron cloud, which is a favorable point for the use of these doped glasses as the core composition in optical fiber production.

\subsection{Absorption and Fluorescence Spectra}
Fig.~\ref{fig1} presents the UV and visible absorption spectra recorded on 0.1, 0.2 and 0.5\% Pr$^{3+}$-doped IZSB glasses samples. The visible absorption bands originating from praseodymium electronic transitions from the fundamental level to the excited ones are highlighted in the inset of Fig. \ref{fig1}. One can notice that the orange emission band under study in this work – located at 604 nm – is quite close to the absorption band centered at 588 nm that corresponds to the electronic transition from the ground state ($^3$H$_4$) to excited level $^1$D$_2$. 

\begin{figure}[h]
      \includegraphics[width=0.5\linewidth]{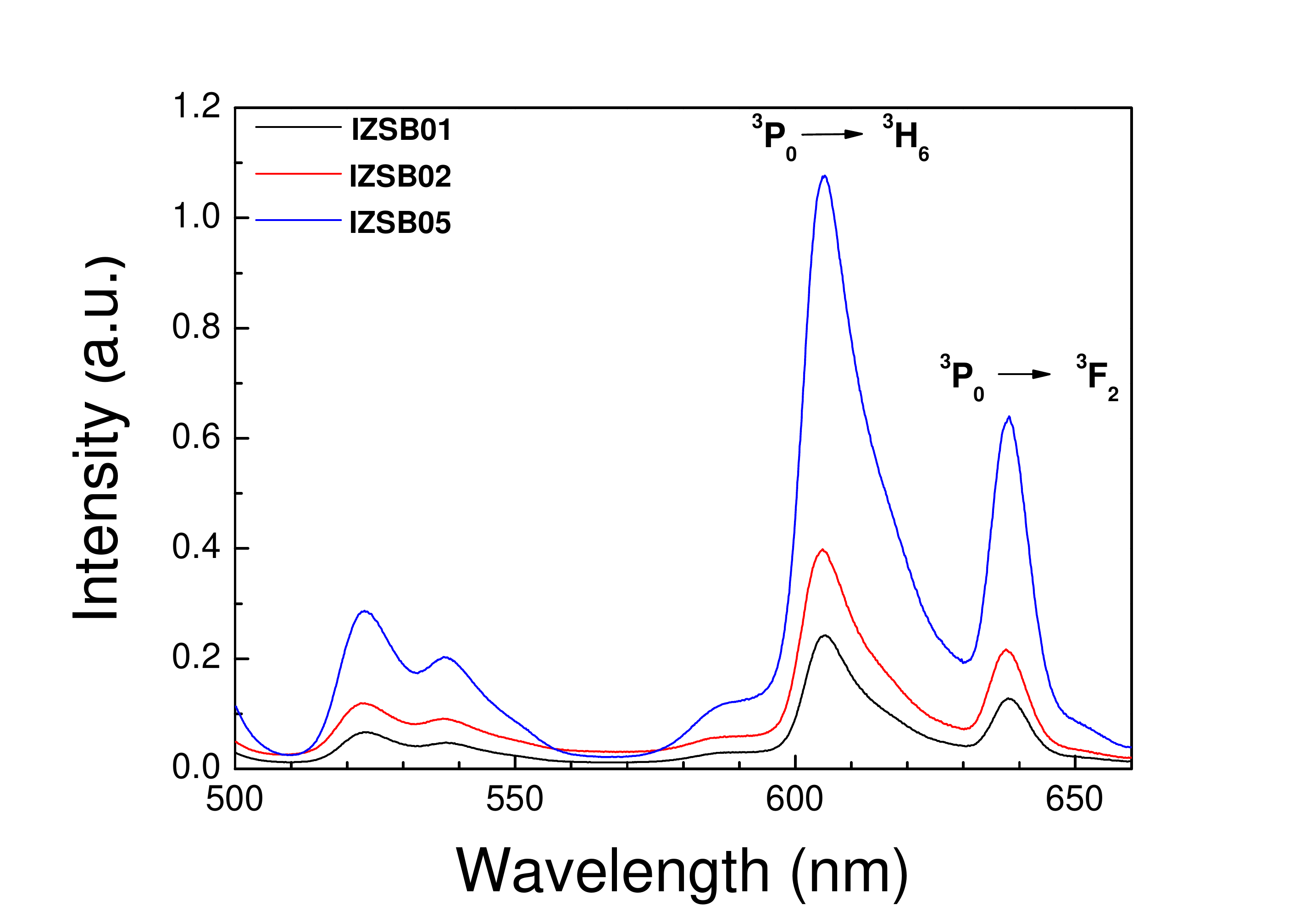}
  \caption{Emission spectra of the Pr$^{3+}$-doped IZSB glasses pumped at 480 nm with a Xe lamp of power of 450~mW.}\label{fig2}
\end{figure}

This is why, in order to accurately measure fluorescence spectra, the excitation beam has been focused close to the surface of the glass samples in order to avoid re-absorption of the emitted photons by the ground state Pr$^{3+}$ ions. In these excitation conditions, the fluorescence emission from the $^1$D$_2$ level was only observed under resonant pumping \cite{Balda2003}. The corresponding fluorescence spectra under blue excitation at 480 nm are reproduced in Fig.~\ref{fig2}.  One can recognize the strong emission lines from level $^3$P$_0$ and  an emission band from levels $^3$P$_1$ and $^1$I$_6$. The typical emission lines of praseodymium ions in the visible can be observed centered at 523, 537, 604 and 637~nm. These lines are associated with the follow transitions: $^3$P$_1/^1$I$_6 \rightarrow ^3$H$_5$, $^3$P$_0 \rightarrow ^3$H$_5$, $^3$P$_0 \rightarrow ^3$H$_6$, and $^3$P$_0 \rightarrow ^3$F$_2$, respectively \cite{Yang2006}. For our application, it is worth noticing the relatively broad ($\sim 15$ nm full width at half maximum), quite intense band around 604 nm in the emission spectrum. This should allow to reach laser emission at 606~nm, which is the excitation wavelength of the $^3$H$_4 \to^1$D$_2$ transition of Pr$^{3+}$ in Y$_2$SiO$_5$, the most popular ion/matrix combination for quantum information processing in rare earth doped materials \cite{Longdell2005,Rippe2005}.

\begin{figure}[h]
      \includegraphics[width=0.4\linewidth]{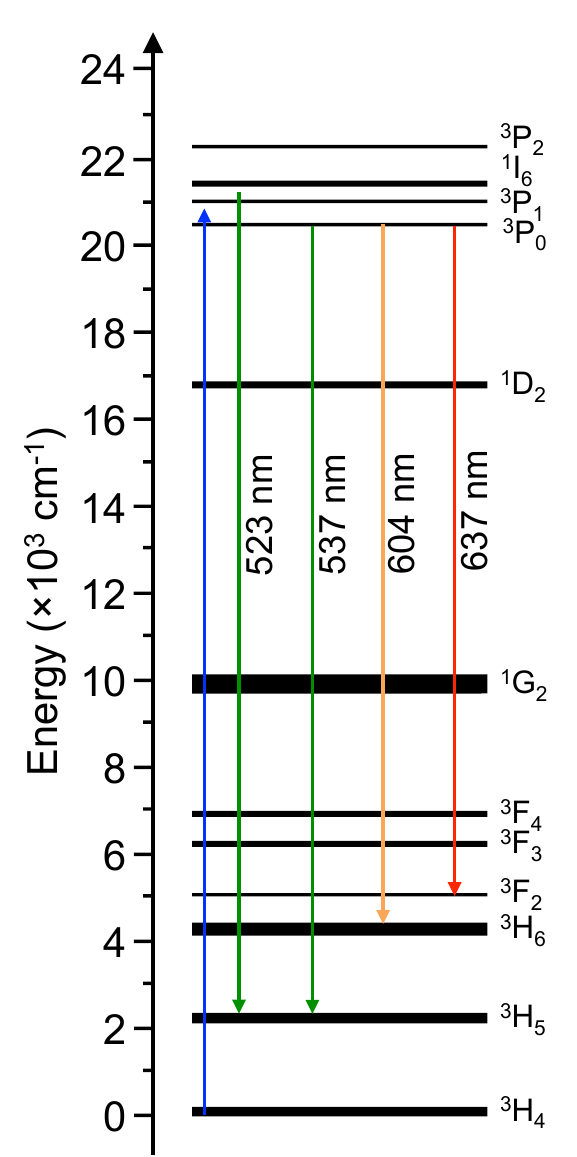}
  \caption{Energy level diagram of Pr$^{3+}$ ion with the proposed down conversion mechanism in the IZSB glass matrix upon 480~nm Xe lamp excitation.}\label{fig3}
\end{figure}

The different transitions are easier to follow on Fig.~\ref{fig3}, which presents the energy level diagram of Pr$^{3+}$ ions in IZSB glasses with the radiative electronic transitions corresponding to the emission bands studied in the present work. The photoluminescence excitation at 480~nm is due to excitation from $^3$H$_4$ to $^3$P$_0$ excited state in the 4f$^2$ configuration of Pr$^{3+}$. Such excitation transitions in Pr$^{3+}$ doped fluoride glasses have already been reported in other hosts \cite{Inoue2003,Nazabal2012}. 

\subsection{$^3$P$_0$ Lifetime Measurements}
\begin{figure}[h]
      \includegraphics[width=0.5\linewidth]{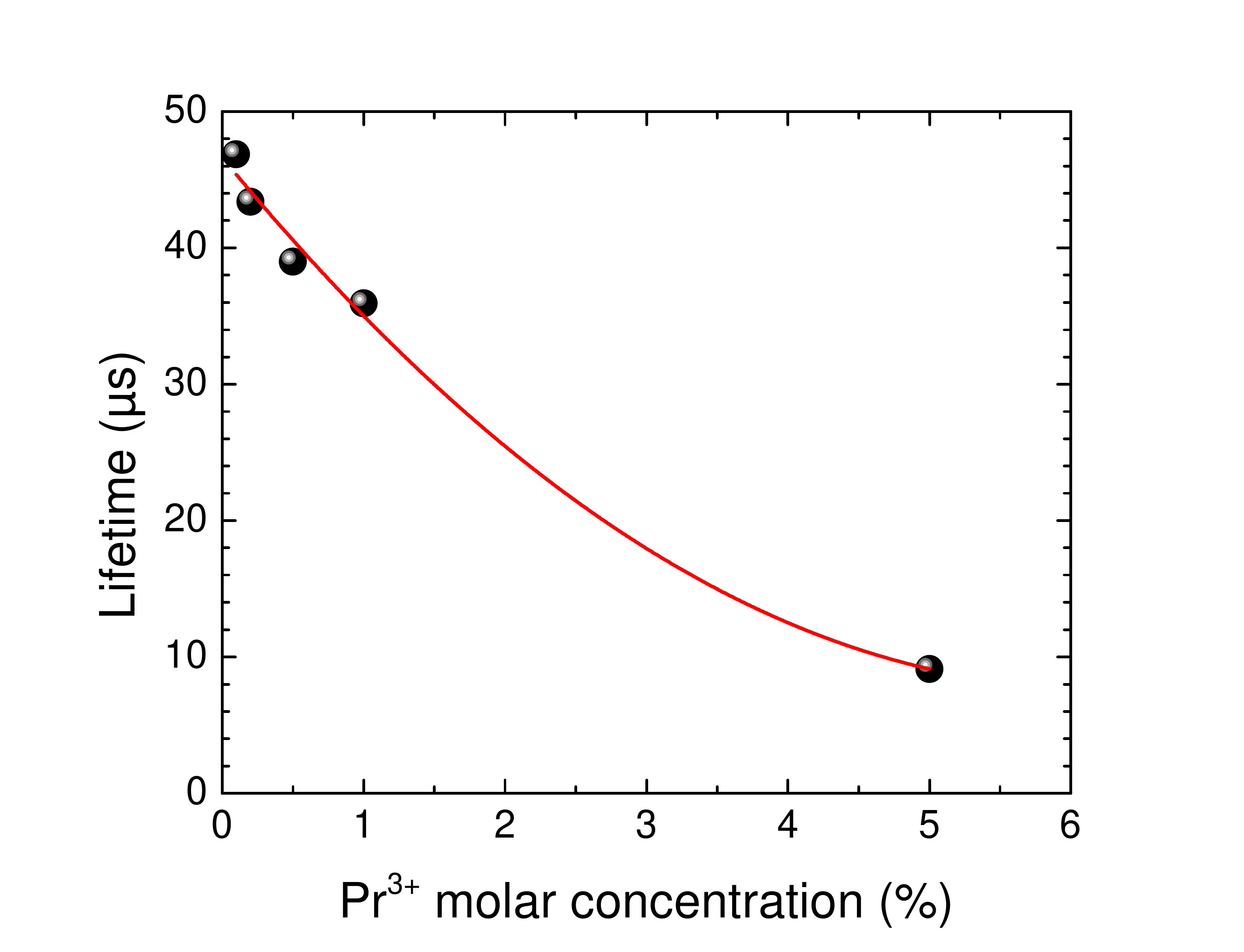}
  \caption{Evolution of the lifetime of the $^3$P$_0$ level versus Pr$^{3+}$ ion concentration, measured using the fluorescence according to the $^3$P$_0\rightarrow^3$P$_6$ transition. The red line is just a guide to the eye.}\label{fig4}
\end{figure}

An important parameter for potential laser operation of our IZSB glasses is the lifetime of the upper level of the transition, i. e., of the $^3$P$_0$ level. We measured the evolution of the lifetime of this level by recording the fluorescence decay along the $^3$P$_0\rightarrow^3$H$_6$ transition for pulsed excitation at 480~nm. The results are summarized in Fig.~\ref{fig4}. One can see that concentration quenching effect remain relatively modest for concentrations up to 1 mol.\%.

\begin{figure}[h!]
      \includegraphics[width=0.5\linewidth]{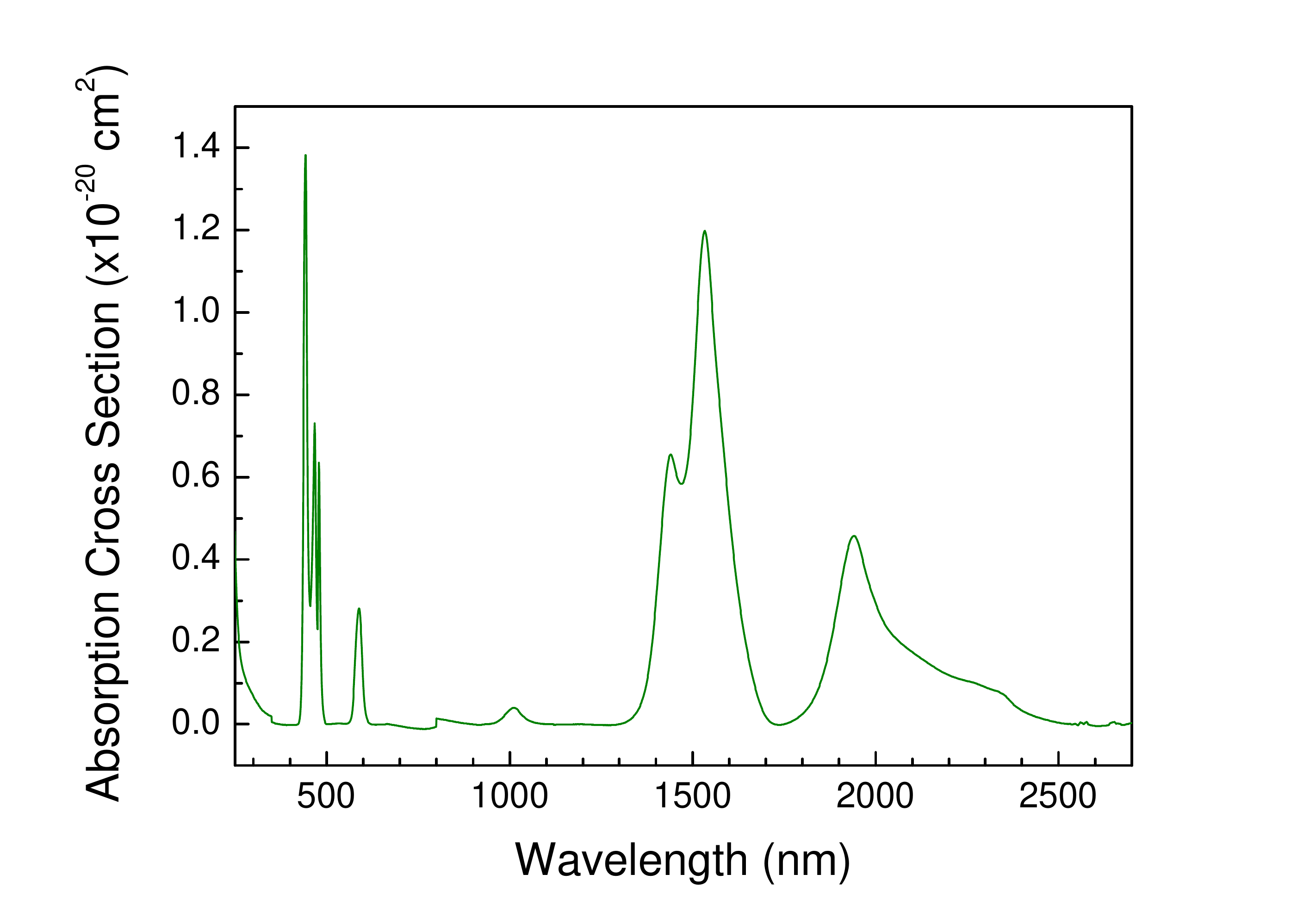}
  \caption{Absorption spectrum of sample IZSB5 used for the Judd-Ofelt analysis.}\label{fig5}
\end{figure}

\subsection{Judd-Ofelt modeling}

\begin{table*}[]
\caption{Experimental ($f_\mathrm{exp}$) and calculated ($f_\mathrm{cal}$) oscillator strengths and corresponding Judd-Ofelt parameters for IZSB5 glass (5 \% Pr$^{3+}$ doping level). All transitions start from the 
$^3\mathrm{H}_4$ multiplet. }
\begin{center}
\begin{tabular}{c c c c c c}
Final level & Transition  & $f_\mathrm{exp} \times 10^6$ & $f_\mathrm{cal} \times 10^6$&  \multicolumn{2}{c}{JO parameters} \\
&energy (cm$^{-1}$)&&&\multicolumn{2}{c}{$\times 10^{20}$ (cm$^2$)}\\
 \hline
$^3\mathrm{P}_2$ & 22635 & 9.37  & 3.2 & $\Omega_2$ & 0.7\\
$^3\mathrm{P}_1+^1$I$_6$ & 21670 & 5.51 & 4.4 & $\Omega_4$ & 5.3\\
$^3\mathrm{P}_0$ & 20849 & 2.48 &  3.0 & $\Omega_6$ & 5.0\\
$^1\mathrm{D}_2$ & 17004 & 2.04 & 0.92 \\
$^1\mathrm{G}_4$ & 9889 & 0.256 & 0.13 \\
$^3\mathrm{F}_4$ & 6695 & 2.03 & 3.0 \\
$^3\mathrm{F}_3$ & 6498 & 6.62 & 5.7 \\
$^3\mathrm{F}_2+^3$H$_6$ & 4782 & 2.94 & 3.1\\ \hline
\end{tabular}
\end{center}
\label{JO}
\end{table*}%

In order to gain more insight in the spectroscopic properties of Pr$^{3+}$ embedded in IZSB glasses, we performed Judd-Ofelt modeling of the absorption spectrum of sample IZSB5 reproduced in Fig.~\ref{fig5}.

Experimental oscillator strengths $f_\mathrm{exp}$ were deduced from the absorption 
spectrum of the IZSB5 sample to get reliable values on  weak transitions like $^3$H$
_4\to^1$G$_4$. The values are gathered in Table \ref{JO}. Judd-Ofelt (JO) parameters \cite{Judd1962,Ofelt1962} $\Omega_2$, $\Omega_4$ and $\Omega_6$  were determined from them using a normalized 
method \cite{Goldner1996,Goldner2003}, which allows fitting of relative values instead of absolute ones. As a result, including 
in the fit the $^3$H$_4\to^3$P$_2$ transition, which has a large oscillator 
strength, does not dramatically perturb the JO parameters (significant  changes, negative 
values for $\Omega_2$), which is often the case using the standard method \cite{Peacock1975,Quimby1994}. The 
parameters are given in Table \ref{JO} and are similar to those found in other fluorozirconate 
glasses \cite{Goldner1996}. Thanks to the normalized method, $\Omega_2$ is positive, as required by theory. It 
should be however noted that the agreement between calculated  and experimental oscillator 
strengths is limited and deviations as large as 50 \% are observed (Table \ref{JO}). Assuming 
an error of $\pm$ 5\% on   $f_\mathrm{exp}$, we found RMS$_\mathrm{norm}$ = 20.3 
whereas it was 16 or 18 in other fluorozirconate glasses \cite{Goldner1996}, denoting a less accurate modeling 
in our case. The 38 $\mu$s radiative lifetime of $^3\mathrm{P}_0$ level was calculated  from the JO parameters. The experimental value can be estimated around 50 $\mu$s  from Fig. 4 and this discrepancy is simply attributed to the inaccuracy of the JO modeling in this glass. It suggests however that the lifetimes measured at low Pr$^{3+}$ concentration are essentially radiative, although further experiments should be performed to confirm this conclusion. The calculated oscillator strength of the $^3\mathrm{P}_0\to^3$H$_6$ transition is $8.7\times 10^{-6}$, a rather high value compared to the absorption transition ones. This should be favorable to orange laser emission.

\section{Conclusion}
In conclusion, we have synthesized praseodymium doped fluoroindate glasses with good optical quality. Strong orange emission centered at 604~nm with a full width at half maximum of 15~nm has been observed. Moreover, the characteristics of the measured absorption spectra show that diode pumping at wavelengths close to 445~nm is conceivable. At praseodymium concentrations lower than 0.5\,\%, the $^3$P$_0$ level was found to exhibit a lifetime longer than 40~$\mu$s, which is compatible with efficient laser emission. 

Further studies could also consist in elaborating such glasses under the form of preforms for optical fibers, or in trying surface ion exchange processes to locally increase the refractive index in order to build active waveguides \cite{Olivier2012}.

\section*{Acknowledgments}
The authors are happy to thank Patrick Aschehoug for technical support. The Funda\c c\~ao de Amparo \`a Pesquisa do Estado de S\~ao Paulo (FAPESP) and the CNPq (Conselho Nacional de Desenvolvimento Cient\'ifico e Tecnol\'ogico) are gratefully acknowledged for their financial support. This work was performed in the framework of the CNRS/FAPESP project No. 24458. 





\bibliographystyle{model1-num-names}
\bibliography{Fluoroindates}







\end{document}